\documentstyle[11pt,epsf]{article}
\input epsf
\begin{document}
\baselineskip 100pt
\renewcommand{\baselinestretch}{1.5}
\renewcommand{\arraystretch}{0.666666666}
{\large
\parskip.2in
\newcommand{\be}{\begin{equation}}
\newcommand{\ee}{\end{equation}}
\newcommand{\br}{\bar}
\newcommand{\fr}{\frac}
\newcommand{\lm}{\lambda}
\newcommand{\ra}{\rightarrow}
\newcommand{\al}{\alpha}
\newcommand{\bt}{\beta}
\newcommand{\pr}{\partial}
\newcommand{\hs}{\hspace{5mm}}
\newcommand{\up}{\upsilon}
\newcommand{\dg}{\dagger}
\newcommand{\vphi}{\vec{\phi}}
\newcommand{\ve}{\varepsilon}
\newcommand{\acc}{\\[3mm]}
\newcommand{\dl}{\delta} 
\newcommand{\ie}{{\it i.e.}}
\newcommand{\cmod}[1]{ \vert #1 \vert ^2 }
\newcommand{\mod}[1]{ \vert #1 \vert }
\newcommand{\nhat}{\mbox{\boldmath$\hat n$}}
\newcommand{\C}{{\rm C}}

\title{\hbox{\hspace{18mm}{\Large{\bf Low Energy States in the $SU(N)$
Skyrme Models}}
\raisebox{16mm}{\hspace{-25mm}{\large DTP98/51\hs UKC98/42}}}}

\author{
T. Ioannidou\thanks{e-mail address:
T.Ioannidou@ukc.ac.uk},\,
\\
Institute of Mathematics, University of Kent at Canterbury, \\
Canterbury CT2 7NF, UK
\acc
\acc
B. Piette\thanks{e-mail address: B.M.A.G.Piette@durham.ac.uk} \,
and
W. J. Zakrzewski\thanks{e-mail address:W.J.Zakrzewski@durham.ac.uk}
\\
Department of Mathematical Sciences,University of Durham, \\
Durham DH1 3LE, UK\\} \date{}

\maketitle
\begin{abstract}
We show that any solution of the $SU(2)$ Skyrme model can be used to give
a topologically trivial solution of the $SU(4)$ one. 
In addition, we extend the method introduced in 
\cite{HMS} and use harmonic maps from $S^2$ to $CP^{(N-1)}$ to construct 
low energy configurations of the $SU(N)$ Skyrme models.
We show that one of such maps gives an exact, topologically
trivial, solution of the $SU(3)$ model. 
We study various properties of these maps and show that, in general,
their energies are only
marginally higher than the energies of the corresponding $SU(2)$
embeddings.
Moreover, we show that the baryon (and energy) densities of the $SU(3)$
configurations with baryon number $B = 2-4$ are more symmetrical than 
their $SU(2)$ analogues. We also present the baryon densities for
the $B=5$ and $B=6$ configurations and discuss their symmetries.
\end{abstract}

\section{INTRODUCTION}

The Skyrme model is now well established as an effective classical
theory used to
describe nuclei \cite{Skyr1,Wit} for which the field, which describes  
pions, is valued in $SU(N)$.
To have finite energy configurations, one must require that the field
$U(\vec{ x},t)$ goes to a constant matrix, say $I$, at
spatial 
infinity:
$U \ra I$ as $|\vec{x}| \ra \infty$. 
This effectively compactifies the three dimensional Euclidean space into
$S^3$ and hence implies that the field configurations of the Skyrme model
can be considered as mappings from $S^3$ into $SU(N)$.

In terms of the algebra valued currents $\pr_\mu U\, U^{-1}$,
the Lagrangian is,
\be
{\cal L}=\fr{F^2}{16}\,\mbox{tr}\left(\pr_\mu U\,\pr^\mu U^{-1}\right)
+\fr{1}{32 a^2}\, \mbox{tr}\left[\pr_\mu U\,
U^{-1},\pr_\nu U\, U^{-1}\right]^2,
\label{lag}
\ee
where  $U(\vec x,t)$ is  an $SU(N)$ valued scalar field, 
$F\approx 189$ MeV is the pion decay constant and $a$ is a dimensional
constant. 
              
The last term, called the Skyrme term,  stabilises the
solitons and, in addition, introduces small interaction forces
between them. 
Their nature depends on the relative orientation of skyrmions in their
internal space.
If we want to study interactions of physical mesons we have to
introduce further terms which are responsible for the meson masses, {\it 
ie }  terms of the form
\be
{\cal L}_m=\fr{F^2}{16}\,M_\pi^2\,\mbox{tr}\left(U^{-1}+U-2\,{\it I}\right).
\label{mass}
\ee
Such terms play a more significant role in lower dimensions since in
(2+1) dimensions their presence, together with the Skyrme term, is
required to stabilise the solitons.
However, in (3+1) dimensions  the solitons are stable even when 
$M_{\pi} = 0$.
In what follows, we will set $M_{\pi}=0$ except in section 5.

In this paper, we concentrate our attention on studying the static 
properties of the model and we  consider $U(\vec x)$
fields,  which are stationary points of the
energy functional. 
After rescaling the space-time coordinates \cite{M,BS} 
$ \vec{x} \rightarrow {2 \vec{x}/ a F}$ and defining 
$m_{\pi} =  M_{\pi} {2 / a F}$, we can write the static energy corresponding 
to the lagrangian (\ref{lag}) and (\ref{mass}) as
\be
E={F\over 4 a} \int_{R^3}\left\{-\fr{1}{2}\,\mbox{tr}\left(\pr_iU\,
U^{-1}\right)^2-\fr{1}{16}\,\mbox{tr}\left[\pr_iU\,
U^{-1},\pr_j U\, U^{-1}\right]^2 -
{m_{\pi}^2\over 2}\mbox{tr}(U^{-1}+U-2\,{\it I})\right\}d^3\vec x.
\label{gene}
\ee
From now on, we will take $F/ 4 a = 1 /12 \pi^2$ so that the energy is 
expressed in the same units as the baryon number. 
In this case, the static fields $U$ obey the equation 
\be
\pr_i\left(\pr_iU\, U^{-1}-\fr{1}{4}\,[\pr_jU\, U^{-1},[\pr_j U\, U^{-1},
\pr_iU\, U^{-1}]]\right)-\fr{m_\pi^2}{4}\left(U-U^{-1}\right)=0.
\label{geq}
\ee
As the third homotopy class of $SU(N)$ is $Z$ every field configuration is
characterised by an integer:
\be
B=\fr{1}{24\pi^2}\int_{R^3} \ve_{ijk}\,\mbox{tr}\left(\pr_i U\,
U^{-1}\pr_j U\, U^{-1}
\pr_k U\, U^{-1}\right)d^3\vec x,
\ee
which is to be interpreted as the baryon number 
\cite{Wit,Skyr}; therefore, the lowest energy state in the $B=1$ sector
can be identified with the (classical) nucleon.

So far most of the studies involving the Skyrme model have concentrated on
the $SU(2)$ version of the model and its embeddings into $SU(N)$. 
The simplest nontrivial classical solution involves a single skyrmion
($B=1$) and has already been discussed by Skyrme \cite{Skyr1}.
The energy density of this solution is radially symmetric
 and, as a result, using the so-called
hedgehog ansatz one can reduce (\ref{geq}) to an ordinary differential
equation which can then be solved numerically.

Many solutions with $B > 1$ have also been computed numerically
and in all cases the solutions are very symmetrical (cf. Battye 
et al. \cite{BS} and references therein).  
The energy density of the two skyrmion solution forms a torus,
while the energy density of the $B=3$ solution has the symmetry of a
tetrahedron.
For larger $B$ the solutions describe  semi-radially symmetric structures
in which skyrmions break up into connected parts which are all located
on a spherical hollow shell and, as was shown in \cite{HMS}, the
positions  of these skyrmionic parts on $S^2$ are very symmetrical.

Recently, Houghton et al. \cite{HMS} have showed that by using rational 
maps from $S^2$ to $S^2$ one can easily construct field configurations for the
$SU(2)$ model which are close to being solutions of the model: they have
energies slightly 
higher than the energies of the exact solutions found numerically
but the symmetries of the baryon and energy densities
are the same. 
When these configurations are used as initial conditions in a relaxation 
program, the fields do not change much as they evolve towards the exact
solutions.

All this work has involved the $SU(2)$ skyrmions; however, so far, very
little has been done for the $SU(N)$ model when $N > 2 $. 
An interesting  question then arises as to whether there are any finite energy 
solutions of the $SU(N)$ ($N > 2$) model which are not embeddings of the 
$SU(2)$ model and, if they exist, whether they have lower
energies than their $SU(2)$ counterparts.

The first example of such a {\it non-embedding} configuration for a higher
group was the $SO(3)$ soliton, which corresponds to a bound system of
two skyrmions, and which was found using the chiral field ansatz 
by Balachandran et al. \cite{BBLRS}. 
Another configuration, with a large $SU(3)$ strangeness content,
was found by Kopeliovich et al. \cite{Kop}.
However, all  other known skyrmion configurations seem to have been 
the embeddings of the solutions of the $SU(2)$ model.

\section{$SU(2)$ EMBEDDINGS}

As we have said before, any solution of the $SU(N^\prime)$ model is
automatically a solution of any $SU(N)$ model as long as
$N^\prime<N$; simply by completing the entries of the larger matrix with
1's along the diagonal and 0's off diagonal. 
The energy, the baryon number and all other
properties are unchanged by this operation and so such embeddings
have the same properties as the original fields.

However, there exist further, less obvious embeddings in which new fields
have different properties from the original ones. 
In particular, one can show that any solution of the $SU(2)$ model
generates a solution of the $SU(4)$ one. 

A special feature of the $SU(2)$ field is that it can be written as
\be
U=\vphi\,\vec{\tau},
\ee
where $\vec{\tau}=(1,i \sigma_1,i\sigma_2,i\sigma_3)$ and the $\sigma$'s
stand for the Pauli matrices.
The unitarity of $U$ requires that $\vec {\phi}\cdot \vec{\phi}=1$; and
 the energy density of this $SU(2)$ field is 
\be
{\cal E}_2=\,(\partial_i\vphi\cdot\partial_i\vphi)
+\frac{1}{2}\left[(\partial_i\vphi
\cdot\partial_i\vphi)\sp2-(\partial_i\vphi\cdot\partial_j\vphi)\sp2\right].
\label{en}
\ee
Moreover, the equations of motion which follow from (\ref{geq})
are
\begin{eqnarray}
\vphi_{ii}+ \vphi_{ii}\,(\vphi_j\cdot \vphi_j)+
\,\vphi_i\,(\vphi_{ij}\cdot \vphi_j)
-\vphi_{ij}\,(\vphi_i\cdot\vphi_j)-\vphi_j\,(\vphi_{ii}\cdot
\vphi_j)-\vphi\,(\vphi\cdot \vphi_{ii})&&\nonumber \\
-\vphi\left[(\vphi\cdot\vphi_{ii})\,(\vphi_j\cdot\vphi_j)-
(\vphi\cdot\vphi_{ij})\,
(\vphi_i\cdot\vphi_j)\right]=0,&&
\label{phieq}
\end{eqnarray}
where $\vphi_i=\partial_i\vphi$ and $\vphi_{ij}=\partial_{ij}\vphi$.

Notice now that we can construct an $SU(4)$ field out of any $S^3$ field 
$\vphi$ by taking
\be
U=U_0 \,\left({\it I}-2\,\vec\phi\otimes\vec\phi\,\sp{\dagger}\right),
\label{us3}
\ee 
where $U_0$ is a constant matrix.
Note that $\det U=-\det U_0$ and so by choosing
$U_0$ to be a constant matrix of determinant -1 we see that $U$ is unitary.
In this case, 
$U\sp{-1}=({\it I}-2\,\vec\phi\otimes\vec\phi\,\sp{\dagger})\,U_0\sp{-1}$
and the $U_0$'s cancel in (\ref{geq}).

To derive the equation that the field $\vphi$ must satisfy so that (\ref{us3})
is a solution of (\ref{geq}), with $m_{\pi} = 0$, we note that the condition
$\vec\phi\cdot\vec\phi=1$ gives 
\be
\partial_j UU\sp{-1}\,=\,2\,\partial_j\vphi\otimes
\vphi\sp{\dagger}\,-\,2\,\vphi\otimes\partial_j\vphi\sp{\dagger},
\label{derivative}
\ee
and so (\ref{geq}) becomes 
\begin{eqnarray}
\!\!\!\! \!\!\!\!\! \!\!\!\!\!
2\partial_i\left[\vphi\otimes\partial_j\vphi\sp{\dagger}
\left(\partial_j\vphi\cdot \partial_i
\vphi\right)\!-\!\vphi\otimes\partial_i\vphi
\sp{\dagger}\left(\partial_j\vphi\cdot\partial_j\vphi\right)
\!+\!\partial_i\vphi\otimes\vphi\sp{\dagger}
\left(\partial_j\vphi\cdot\partial_j
\vphi\right)\!-\!\partial_j\vphi\otimes\vphi\sp{\dagger}
\left(\partial_i\vphi\cdot
\partial_j\vphi\right)\right]&&\nonumber\\
\hspace{101mm} +\partial_i\left(\partial_i\vphi\otimes\vphi\sp{\dagger}
-\vphi\otimes\partial_i\vphi\sp{\dagger}
\right)=0.&&\nonumber\\
\label{eqaa}
\end{eqnarray}
One can easily show that the equations (\ref{eqaa}) and (\ref{phieq}) are
equivalent, implying that $\vphi$ satisfies the equation of the $SU(2)$ 
Skyrme model (\ref{eqaa}). We can thus conclude that any solution of the 
$SU(2)$ model can be transformed into a solution of the $SU(4)$ model 
by the embedding (\ref{us3}). 

The solutions obtained this way are topologically trivial since their baryon
density vanishes identically.
Moreover, their energy density is
\be
{\cal E}_3=4\,(\partial_i\vphi\cdot\partial_i\vphi)+2\left[(\partial_i\vphi
\cdot\partial_i\vphi)\sp2-(\partial_i\vphi\cdot\partial_j\vphi)\sp2\right],
\ee
{\it ie} it is four times larger than the corresponding energy of the original 
$SU(2)$ field (\ref{en}).

This suggests that these particular $SU(4)$ solutions may be interpreted as
states corresponding to $2B$ skyrmions and $2B$
anti-skyrmions, where $B$ is the baryon number of the original $SU(2)$ 
solution.
Incidentally, a similar situation arises in 2-dimensions
where any $B$ solitonic solution of the $CP\sp1$ model gives a topologically
trivial solution of the $CP\sp2$ model which can be interpreted as
a bound state of $2B$ solitons and $2B$ anti-solitons \cite{Zak}.

\section{HARMONIC MAPS ANSATZ}

Recently, Houghton et al. \cite{HMS} exploited the similarity of the
energy densities of the multi-skyrmion solutions to those of the $SU(2)$ BPS
monopoles and presented a new ansatz for
constructing $SU(2)$ multi-skyrmion fields; 
based on rational maps of the two dimensional sphere $S^2$. 
Namely, they showed that  solutions of the Skyrme model can be well
approximated by the expressions of the form 
\be
U(r,\theta,\phi) = \exp(i g(r)\, \mbox{\boldmath $\hat n \cdot \sigma$}),
\label{uhsu2}
\ee
where $(r, \theta, \phi)$ are the usual polar coordinates on $\bf R^3$, and
\be
\nhat = {1\over 1 + \cmod{R}}(2\Re(R), 2\Im(R),1-\cmod{R}),
\ee
where $R$ are rational functions of $\xi=\tan(\theta/2) e^{i \phi}$ and 
where $g(r)$ is a real function satisfying the boundary conditions: $g(0)
= \pi$ and $g(\infty) = 0$.
In other words, the configuration (\ref{uhsu2}) involves a radial profile
function $g(r)$ and a rational map from the
two dimensional sphere of radius $r$ which can be identified with a sphere
centered at the origin, in ${\bf R^3}$, into a $S^2$ submanifold of
$SU(2) \equiv S^3$. 
Moreover, it is easy to  check that the baryon number $B$ is given by the
degree of the rational map \nhat.

To determine $g$ and \nhat\ one must insert (\ref{uhsu2}) into 
(\ref{gene}) and minimise the energy. It turns out that, for this minimum, 
\nhat\ must be a rational map with a large discrete symmetry 
and that $g(r)$ satisfies an ordinary differential equation.

In this section we show that this idea of Houghton et al. can be
generalised to  $SU(N)$.
Using the polar coordinates $(r, \theta, \phi)$ in ${\bf R^3}$, our 
generalisation of Houghton et al.'s ansatz is to consider $U$ of the form 
\begin{eqnarray}
U(r,\theta, \phi)&=&e^{2ig(r)(P-{\it I}/N)}\nonumber\\
        &=&e^{-2ig(r)/N}\left({\it I}+(e^{2ig}-1)P\right),
\label{uni}
\end{eqnarray}
where $P$ is a $N\times N$ hermitian projector which depends only on the
angular variables $(\theta,\phi)$ and $g(r)$ is the radial
profile function.
Note that, the matrix $P$ can be thought of as a mapping from $S^2$ into 
$CP^{(N-1)}$.
Hence it is convenient, rather than using the polar coordinates, to
map the sphere onto the complex plane via a stereographic projection and, 
instead of $\theta$ and $\phi$, use the complex coordinate $\xi$ and its
conjugate. Thus, $P$ can be written as
\be
P(V)=\fr{V \otimes V^\dg}{|V|^2},
\label{for}
\ee
where $V$ is a $N$ component complex vector (dependent on $\xi$ and
$\bar{\xi}$).

For (\ref{uni}) to be well-defined at the origin, like (\ref{uhsu2}),
the radial profile function $g(r)$ has to satisfy
$g(0)=\pi$ while the boundary value $U \ra I$ at $r=\infty$ requires that 
$g(\infty)=0$.
An attractive feature of the ansatz (\ref{uni}) is that it leads to a
simple expression for the energy density which can  be successively minimized
with respect to the parameters of the projector $P$ and then with respect
to the shape of the profile function $g(r)$. This is then expected to give good 
approximations to multi-skyrmion field configurations.

Moreover,  we will show that this method not only allows us
to find such field configurations but also 
gives us an exact non-topological solution of the $SU(3)$ Skyrme model.
We will also present some upper bounds on the energy of some
multi-skyrmion field configurations in the $SU(N)$ model (with radially
symmetric energy density distribution). In what follows, we restrict our attention to the case $m_{\pi} = 0$.

To find the exact solution of the $SU(3)$ model we put (\ref{uni}) into
 (\ref{geq}) and obtain
\begin{eqnarray}
&&\!\!\!\!\!\!\!\!-\fr{2i}{r^2}\,\pr_r(g_r\,r^2)\,
(\fr{{\it I}}{N}-P)-\fr{i}{2r^2}\,\pr_r
(g_r|A|^2)\,(1+|\xi|^2)^2
\left([P_{\bar{\xi}},P\,P_{\xi}]+[P_{\xi},P\,P_{\bar{\xi}}]\right)\!
\nonumber\\
&&\!\!\!\!\!\!\!\!+\fr{\bar{A}|A|^2}{16r^4}(1+|\xi|^2)^2\left\{\pr_{
\bar{\xi}}\left
((1+|\xi|^2)^2 [P_{\xi},[P_{\bar{\xi}},P_{\xi}]]\right)\!-\!
\pr_{{\xi}}\left((1+|\xi|^2)^2 
[P_{\bar{\xi}},[P_{\bar{\xi}},P_{\xi}]]\right)
\right\}\!\nonumber\\
&&\!\!\!\!\!\!\!\!+\fr{|A|^4}{16r^4}(1\!+\!|\xi|^2)^2
\left\{\pr_{\bar{\xi}}\left
((1\!+\!|\xi|^2)^2 [P_{\xi}P,[P_{\bar{\xi}},P_{\xi}]]\right)\!-\!
\pr_{{\xi}}\left((1\!+\!|\xi|^2)^2
[P_{\bar{\xi}}P,[P_{\bar{\xi}},P_{\xi}]]\right)
\right\}\nonumber\\
&&\!\!\!\!\!\!\!\!+\fr{1}{r^2}(1+g^2_r)(1+|\xi|^2)^2\left[\bar{A}\,
P_{\xi\bar{\xi}}
+\fr{|A|^2}{2}\left(\pr_\xi(P_{\bar{\xi}}\,P)
+\pr_{\bar{\xi}}(P_{\xi}\,P)\right)\right]=0,
\label{peq}
\end{eqnarray}
where $A=e^{-2ig}-1$.

Moreover, the  energy (\ref{gene}) simplifies to
\be
E={1 \over 3 \pi} 
\int dr \left(A_N\, g_r^2 \,r^2+2{\cal N}\, \sin^2g\, (1+g_r^2)
+{\cal I}\, \fr{\sin^4 g}{r^2}\right),
\label{ene}	
\ee
where
\begin{eqnarray}
\label{al}
A_N&=& \fr{2}{N}(N-1),\\
\label{NN}{\cal N}&=&\fr{i}{2\pi}\int
d\xi\,d\bar{\xi}\,\mbox{tr}\left(|\pr_\xi P|^2\right),\\
\label{II}{\cal I}&=&\fr{i}{4\pi}\int d\xi \,d\bar{\xi}\, (1+|\xi|^2)^2
\,\mbox{tr}\left([\pr_\xi P,\,\pr_{\bar{\xi}} P]\sp2\right).
\end{eqnarray}

As the  integrals ${\cal N}$ and ${\cal I}$ in (\ref{ene}) are
independent of $r$, we can minimise (\ref{ene}) by first minimising
${\cal N}$ and ${\cal I}$ as functions of $P$  and then with respect to
the profile function $g$. 

However, since ${\cal N}$ is the expression for the
energy of the two dimensional Euclidean $CP^2$ sigma model, all classical
solutions contain  the so-called self-dual solutions,
instantons or holomorphic maps from $S^2$ into $CP^{(N-1)}$, first 
given in \cite{Dada}, which are given by the projector $P$ of the form 
(\ref{for}) 
with $V=f(\xi)$. In this case, the energy $\cal{N}$ is given by the degree of 
$f$, {\it ie} the degree of the highest order polynomial in $\xi$ 
among the components of $f$ after all their common
factors  have been canceled.

By a Bogomolny-type argument it can be shown that
\begin{eqnarray}
E&&\!\!\!\!\!\!\!\!={1\over 3\pi}\!\! \int\! dr\left[\left(g_r
r\sqrt{A_N}+\sqrt{{\cal
I}}\,\fr{\sin^2
g}{r}\right)^2
+2N\sin^2 g \,(1+g_r)^2-2g_r\, \sin^2 g \left(2N+\sqrt{A_N\, {\cal
I}}\right)\,\right]
\nonumber \\
&&\!\!\!\!\!\!\!\!\geq {1 \over 3}\,\left(2\,N +\sqrt{A_N\,{\cal
I}}\,\,\right).
\end{eqnarray}

Finally, the baryon number for this ansatz is given by
\begin{eqnarray}
B&=&\fr{i}{\pi^2}\int
d\xi\,d\bar{\xi}\,\mbox{tr}\left(P\,[\pr_\xi
P,\,\pr_{\bar{\xi}} P]\right)
\int_{0}^{\infty}dr\, \sin^2g \, g_r\,\label{Q1}
\nonumber\\
&=&\fr{i}{2\pi}\int
d\xi\,d\bar{\xi}\,\mbox{tr}\left(P\,[\pr_{\bar{\xi}}
P,\,\pr_{\xi} P]\right),
\label{barQ}
\end{eqnarray}
which is the topological charge of the two-dimensional $CP^{(N-1)}$
sigma model.

In the next two sections we will show that this ansatz gives us
interesting low energy field configurations of the  $SU(N)$  Skyrme model 
which are not the  $SU(2)$ embeddings. 
To minimise $E$ we will, first of all, fix the baryon number 
${\cal N} = B$ of the configurations we are interested in. We will then 
minimise ${\cal I}$ over all maps of degree $B$ and
then derive a second order differential equation for $g$ by
minimising the energy (\ref{ene}) treating ${\cal N}$ and ${\cal I}$ as 
parameters.

\section{$SU(3)$ EXACT SOLUTION}

When the projector $P$ is analytic, {\it ie} is of the form 
\be
P_0=P(f)=\fr{f\, f^\dg}{|f|^2},
\label{P0}
\ee
where $f$ is a holomorphic vector ({\it ie} whose entries are functions
only of $\xi$) then it satisfies the equation
\be
P_0\,\pr_\xi P_0=0, \hs \hs \hs \pr_\xi P_0\,P_0=\pr_\xi P_0,
\label{P0ex}
\ee
{\it ie} the self-dual equations of  the two dimensional $CP^{(N-1)}$ 
sigma models \cite{Zak}.

Following \cite{DinZak}, we define the operator $P_+$  by its action
on any vector $v \in \C^N$ as
\be
P_+ v=\pr_\xi v- \fr{v \,(v^\dg \,\pr_\xi v)}{|v|^2},
\ee
and then define further vectors $P^k_+ v$ by induction: 
$P^k_+ v=P_{+}(P^{k-1}_+ v)$.

To proceed further we note the following useful properties of 
$P^k_+ f$ when $f$ is holomorphic:
\begin{eqnarray}
\label{bbb}
&&(P^k_+ f)^\dg \,P^l_+ f=0, \hs \hs \hs k\neq l,\acc
&&\pr_{\bar{\xi}}\left(P^k_+ f\right)=-P^{k-1}_+ f \fr{|P^k_+
f|^2}{|P^{k-1}_+ f|^2},
\hs \hs
\pr_{\xi}\left(\fr{P^{k-1}_+ f}{|P^{k-1}_+ f|^2}\right)=\fr{P^k_+
f}{|P^{k-1}_+f|^2}.
\label{aaa}
\end{eqnarray}
These properties either follow directly from the definition of the $P_+$ 
operator or are very easy to prove \cite{Zak}.

It is also convenient to define projectors corresponding to the 
family of $P_+\sp{k}f$ vectors as follows: 
\be
P_0=P(f),\hs\hs  P_1=P(P_+ f),\hs \hs \dots, \hs \hs P_i=P(P^i_+ f).
\ee

Taking $P = P_i$, for given $i$, and using the above properties we observe 
that all the terms in (\ref{peq}), except the first one, can be 
gathered into one term if and only if 
\be
\fr{|P_+ f|^2}{|f|^2}\equiv \fr{{\cal K}}{(1+|\xi|^2)^2},
\label{cond}
\ee 
where ${\cal K}$ is a constant.
Moreover, for the $SU(2)$ case, the projectors $P_0$ and $P_1$ satisfy the
relation
\be
P_0+P_1 = {\it I},
\ee
and for $f=(1,\xi)$ all the terms in (\ref{peq}) are proportional to one
common matrix thus giving a
second order differential equation for the profile function $g$.
This means that the Skyrme field (\ref{uni}), in the case
when $g$ satisfies its equation, is an exact
solution of the equation (\ref{peq}). A little thought shows
that this is the well known 
hedgehog solution.

Unfortunately this discussion does not generalise to higher  $SU(N)$ 
groups. However, we note that for the $SU(3)$ model, if we take $P = P_1$ and 
use the fact that $P_0+P_1+P_2={\it I}$, all the matrix 
terms in  equation (\ref{peq}) become proportional to each other
leading to a second order differential equation for the profile 
function, if and only if
\be
\fr{|P_+^2 f|^2}{|P_+
f|^2}+\fr{|P_+f|^2}{|f|^2}\equiv \fr{\tilde{\cal K}}{(1+|\xi^2|)^2},
\label{cond2}
\ee 
where $\tilde{\cal K}$ is a constant.
This last condition is satisfied if 
\be
 f=(1,\,\sqrt{2}\, \xi, \,\xi^2)^t.
\label{fsas}
\ee
Thus, by taking $P=P_1$ for $f$ of the form (\ref{fsas}), and requiring
$g$ to satisfy the  equation 
\be
g_{rr}\left(\fr{1}{3}+2\,\fr{ \sin^2 g}{r^2}\right)+\fr{2}{3}\,\fr{g_r}{r}
+\fr{ \sin 2g}{r^2}\left(g_r^2-1-\fr{ \sin^2 g}{r^2}\right)=0,
\ee
we see that (\ref{uni}) is an exact solution of the $SU(3)$ model.

For this solution, the parameters in the energy density can be evaluated 
analytically; we find
\be
A_N=\fr{4}{3}, \hs \hs {\cal N}=4, \hs \hs {\cal I}=4, 
\ee
and the total energy is $E = 3.861$.

To understand what this solution corresponds to we calculate the 
topological charge of this configuration and find 
\be
B=\fr{i}{\pi}\int d\xi\, d\bar{\xi}\,\left(\fr{|P^2_+ f|^2}{|f|^2}
-\fr{|P_+ f|^2}{|f|^2}\right),
\ee 
which due to the conditions (\ref{bbb}), (\ref{aaa}) and (\ref{cond2}) 
is identically zero.

Although the baryon density is identically zero the solution itself
is nontrivial.
This follows from the fact that the $CP^2$ sigma model harmonic map $P_1$ 
corresponds to a mixture of two solitons and two anti-solitons.
Thus it seems reasonable to interpret this solution 
as describing a bound state of two skyrmions
and two anti-skyrmions and as such to be unstable,
{\it ie} correspond to  a saddle point of the energy. 
However, let us emphasize, 
once again, that this field configuration is a 
genuine solution of the $SU(3)$ Skyrme model.

It is easy to see that this new field configuration has an energy
density distribution shaped like a shell ({\it ie} is radially 
symmetric). To see this note that for this solution, 
$\mbox{tr}\left(|\pr_\xi P|\sp2\right)$ and $\mbox{tr}\left([\pr_\xi
P,\pr_{\bar\xi}P]\sp2\right)$ which appear in (\ref{NN}) and (\ref{II}),
 are proportional to $(1+\vert \xi\vert\sp2)^{-2}$
and $ (1+\vert \xi\vert\sp2)^{-4}$, respectively;
demonstrating this symmetry. 
The radial energy density of this solution is given in Figure 1 and one
sees that it corresponds to a hollow ball.  

\begin{figure}[ht]
\unitlength1cm
\hfil\begin{picture}(6,8)
\epsfxsize=6cm
\epsffile{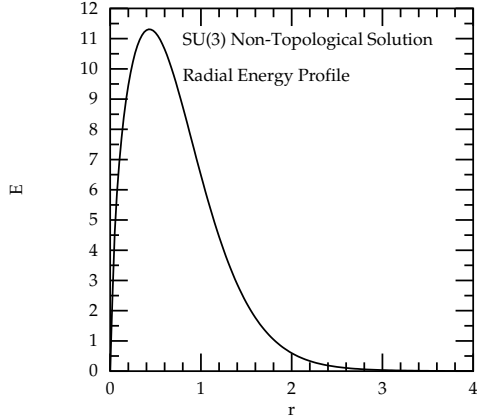}
\end{picture}
\caption{Energy profile of the non-topological solution.}
\end{figure}

\section{APPROXIMATE RADIALLY SYMMETRIC SKYRME FIELDS}

What about further genuine solutions? 
In general, our method does not give us
further solutions but it is a matter of simple 
algebra to show that the condition (\ref{cond}) is true for
any $N \geq 2$ when the modulus of the vector $f$ is some power of
$(1+|\xi|^2)$.
In fact, we have
\begin{eqnarray}
\label{N=2}
N=2, &\hs \hs& f=(1,\xi)^t,\\
\label{N=3}
N=3, && f=(1,\sqrt{2} \xi, \xi^2)^t,\\
\label{N=4}
N=4, && f=(1, \sqrt{3}\xi, \sqrt{3}\xi^2, \xi^3)^t,\\
\label{N=r}
N=n, && f = (f_0, f_1, \dots,  f_{n-1})^t : \hs f_i = \xi^i
\sqrt{C_{i+1}^{n-1}}, 
\end{eqnarray}
where $C_{i+1}^{n-1}$ denotes the binomial coefficients.
Note that in this case, the constant ${\cal K}$ in (\ref{cond}) is
equal to the degree of the vector $f$: {\it ie} ${\cal K} = n$.

Using the condition  (\ref{cond}) the integrals involving $P$ in the energy 
(\ref{ene}) can be evaluated analytically, 
\be
{\cal N}=n, \hs \hs {\cal I}=(N-1)^2=n^2.
\label{ana}
\ee
Using the analyticity of the  projector $P_0$, it is
straightforward to verify that the baryon number of this
field is $B = n$, {\it ie} the degree of $f$.

When $m_{\pi}\ne0$ the energy (\ref{ene}) for the ansatz (\ref{uni}) becomes
\begin{eqnarray}
E\!\!\!&=&\!\!\!{1 \over 3\pi}\!\int\! dr \{A_N\, g_r^2 \,r^2+2{\cal N}\,
\sin^2g\,
(1+g_r^2)+{\cal I}\, \fr{\sin^4 g}{r^2}\nonumber \\
&&\,\hs\hs+m_\pi\sp2\,r\sp2\left[ (N-1)\left(1-\cos(\fr{2\,g}{N})\right)
+1 - \cos\left(\fr{(N-1)\,2g}{N}\right)\right]\}.
\label{enea}	
\end{eqnarray}

Minimising (\ref{enea}) given (\ref{ana})
leads to the following equation for the profile function
\begin{eqnarray}
&&g_{rr}\left(A_N+2n\,\fr{ \sin^2g}{r^2}\right)+2\,A_N\,\fr{g_r}{r}
+\fr{\sin 2g}{r^2}\left(n\,(g_r^2-1)-n^2 \fr{\sin^2g}{r^2}\right)\nonumber\\
&&\hspace{35mm} - \,m\sp2_\pi \left(\fr{N-1}{N}\right)
\left[\sin \left(\fr{2g}{N}\right)+\sin
\left(\fr{(N-1)2g}{N}\right)\right]=0,
\label{telm}
\end{eqnarray}
where $A_N$ is given by (\ref{al}).

Solving (\ref{telm}) to determine $g$ and then 
calculating the energy of the configuration we find that, 
for small $m_{\pi}$, the energy 
for these configurations is a little higher than the energy of the  
$SU(2)$ embedded ansatz with the same baryon number $B$ when the mass is zero. 
However, when the mass increases, the picture changes.

We have looked at field configurations corresponding to
$B=2-4$ for the $SU(2)$ embeddings and for the $SU(N)$ spherical symmetric
fields (\ref{N=3})-(\ref{N=r}) where $N=B+1$ ({\it ie} $SU(3)$ for $B=2$)
and studied the dependence of their energies on $m_\pi$.
In all cases at low values of the mass the embeddings
have lower energies while as the mass increases the energies increase.
However, as the embedding energies increase faster for all low  $B$
there is a value of $m_\pi$ above which the embedding energy is
higher.
Unfortunately, this value of $m_\pi$ is quite large and it increases
with the increase of $B$.

These results are summarised in Table 1, which gives values of the energy 
for different values of the mass, and in Figure 2 where we present the 
dependence on $m_\pi$ of the energies for the embeddings and for the  
radially symmetric fields (\ref{N=3})-(\ref{N=r}).  Note that the energy
per skyrmion of the harmonic ansatz configuration is always lower than the 
energy of a single skyrmion.

\begin{table}[htbp]
\begin{center}
\begin{tabular}{l|l||l|l||l|l||l|l}
    & SU(2) & SU(2) & SU(3) & SU(2) & SU(4) & SU(2) & SU(5)\\
$m_{\pi}$
&$E_{B=1}$&$E_{B=2}$&$E_{B=2}$&$E_{B=3}$&$E_{B=3}$&$E_{B=3}$&$E_{B=4}$ \\
\hline
0   & 1.232 & 2.416 & 2.444 & 3.553 & 3.644 & 4.546 & 4.838\\
0.2 & 1.247 & 2.444 & 2.472 & 3.594 & 3.683 & 4.597 & 4.886\\
1   & 1.416 & 2.795 & 2.808 & 4.125 & 4.172 & 5.270 & 5.520\\
2.23& 1.693 & 3.381 & 3.370 & 5.021 & 5.006 & 6.419 & 6.615\\
7   & 2.510 & 5.101 & 5.030 & 7.634 & 7.478 & 9.776 & 9.880\\
30  & 4.783 & 9.836 & 9.633 &14.793 &14.339 &18.971 &18.948\\
\end{tabular}
\end{center}
\caption{Mass dependence of the energy for the radially symmetric 
configurations in the $SU(2)- SU(5)$ models.}
\end{table}

\begin{figure}[ht]
\unitlength1cm
\hfil\begin{picture}(5,8)
\epsfxsize=6cm
\epsffile{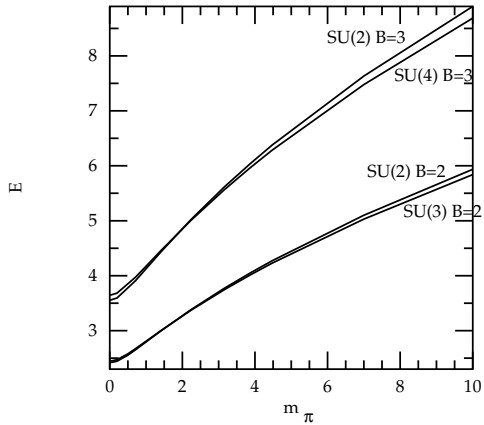}
\end{picture}
\caption{Mass dependence of the $SU(2)$ and $SU(B+1)$ harmonic map 
configurations for $B=2-3$.}
\end{figure}

Our configurations, like the exact solution of the $SU(3)$ model
mentioned above, all have spherically symmetric energy density
distributions ({\it ie} shell like structures).
In Figure 3 we present the curves of the energy density, as a function
of the radius, for the field configurations mentioned above when $m_{\pi} = 0$.
We note that as the topological charge increases (and we consider the $SU(N)$
model with larger $N$)
the effective radius of the distribution also increases.

\begin{figure}[ht]
\unitlength1cm
\hfil\begin{picture}(5,8)
\epsfxsize=6cm
\epsffile{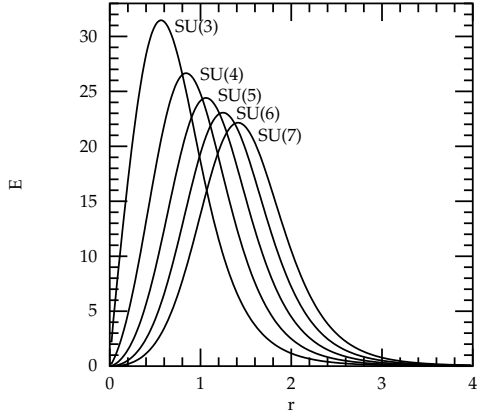}
\end{picture}
\caption{Energy profiles of the $SU(B+1)$ configurations
for $B=2-6$ for $m_{\pi} = 0$.}
\end{figure}

\section{SU(3) CASE}

In this section, we restrict our attention to the $SU(3)$ model, take 
$m_{\pi} = 0$ and construct low energy states with baryon number from one
up to six. 
From now on, $N=3$ and so $A_N$, given by (\ref{al}), becomes 
$A_N=4/3$.

\subsection{GENERAL DISCUSSION}

As in the previous section, we minimise (\ref{ene}) by first 
minimising the  integrals ${\cal N}$ and ${\cal I}$ as functions of $P$ 
and then minimising (\ref{ene}) with respect to the profile function $g$. 

Once again, ${\cal N}$ is minimised by the so-called self-dual 
solutions of the Euclidean $CP^2$ sigma model. They are given by 
(\ref{P0ex}) where $f$ is any polynomial holomorphic vector $f(\xi)$ and 
their energy $\cal{N}$ is given by the degree of $f$.

Next we note  that the angular part of the baryon  charge (\ref{barQ}) 
coincides with the expression for the topological charge of  the $CP^2$ sigma
model and so simplifies to
\be
B =\fr{1}{8\pi}\int dS\, \left(1+|\xi|^2\right)^2\,\fr{|P_+ f|^2}{|f|^2},
\label{bzwg}
\ee
where $dS  \equiv \sin\theta\, d\theta\, d\phi\ = 
2 i\, (1+|\xi|^2)^{-2}\, d\xi\, d\bar{\xi}$.

To minimise (\ref{ene}) for a configuration with a given baryon number $B$, 
we take $f(\xi)$ to be a holomorphic vector of degree $B$ which, by 
construction, minimises ${\cal N}$. 
First we use the global $SU(3)$  invariance of 
the model to reduce the number of parameters to the  moduli space of
the two dimensional sigma model, {\it ie} to
\be 
f=\left(\begin{array}{llcl}
\hs\,\,\,\,\, \,\ \xi^B+a_{B-1} \,\xi^{B-1}+\, \dots \, +a_1\, \xi \\[2mm]
b_{B-1}\,\xi^{B-1}+b_{B-2} \,\xi^{B-2}+\, \dots \,+b_1\, \xi + b_0\\[2mm]
c_{B-2}\,\xi^{B-2}+c_{B-3} \,\xi^{B-3}+\, \dots \,+c_1\, \xi + c_0
\end{array} \right),
\label{genf}
\ee
where all the coefficients are complex except $b_{B-1}$ which can be
taken to be real.
Then we substitute (\ref{P0}) for $f$ of the form (\ref{genf})
into $\cal{I}$ and minimise numerically the integral with respect to all
the coefficients. Finally, treating ${\cal N} = n$ and ${\cal I}$ as two 
fixed parameters, we  minimize (\ref{ene}) by solving the resultant equation 
for $g$:  
\be
g_{rr}\left(1+2{\cal N}\, \fr{3 \sin^2g}{4\, r^2}\right)+
2\,\fr{g_r}{r}+\fr{3\sin 2g}{4\, r^2}\left[{\cal N}(g_r^2-1)-{\cal I}
\,\fr{\sin^2g}{r^2}\right]=0.
\label{eq}
\ee

An interesting feature of the $SU(2)$ multi-skyrmion solutions is the shape
of surfaces of constant energy or baryon density. 
In fact, the energy and the  baryon densities of the skyrmion
solutions look very similar. 
For the baryon density these surfaces look like hollow shell-like
structures with holes in it, while for the energy densities the holes are
partly filled in and so are represented by local minima \cite{BS}. 

In order to investigate the situation for our $SU(3)$ field
configurations, we have to look at the components of $f$ given in
(\ref{genf}) and study their effects on  the density (\ref{bzwg}). 
Writing $f = (K,L,M)^t$ where $K$, $L$ and $M$ are polynomials of
degree $B$, $B-1$ and $B-2$ respectively, the integrand of 
(\ref{Q1}) takes the form
\be
{\cal B} = g_r\, \sin^2 g\,(1+\cmod{\xi})^2\, {\cmod{K_\xi L-L_\xi
K}+\cmod{K_\xi M-M_\xi
K}+\cmod{M_\xi L-L_\xi M} \over \left(
\cmod{K}+\cmod{L}+\cmod{M}\right)^2}.
\label{eeee}
\ee 
Note that the integrand of (\ref{eeee}) is a scalar with respect to $U(3)$ 
transformations applied to the vector $f$. Hence, any modifications of $f$ 
which can be interpreted as such $U(3)$ transformations are symmetries
of (\ref{eeee}).

The radial factor $g_r \sin^2 g$ in (\ref{eeee}) indicates that if the angular
part of the density vanishes, the baryon density will have
radial holes going from the origin to infinity. 
For the density to vanish at some point we must 
require that the three factors in the numerator 
of (\ref{eeee}) must vanish together, {\it
ie}  must have a common root.
This is true, when the three polynomials $R_1 = K_\xi L-L_\xi K$,  $R_2 =
K_\xi M-M_\xi K$  and  $R_3 = M_\xi L-L_\xi K$ have a common factor. 
However, these polynomials have $2 (B-1)$, $2 B -3$ and $2 (B-2)$
roots, respectively; with, in addition, a possible root at infinity ({\it
ie} the south pole of the sphere).
By counting powers we see that the density does not vanish  at 
$\xi=\infty$ unless $L$ is a polynomial of degree less than 
$B-1$.

From this we conclude that the baryon density can have at most $2 B-3 $ holes 
but, in general, it is likely to have fewer holes if any. 
Of course, when some terms in (\ref{eeee}) vanish, the expression may (but
does not have to) have a local minimum. 
Note that this is in complete contrast with  the $SU(2)$ configurations of
Houghton et al. \cite{HMS}  which always have $2 (B-1)$ holes. 
In the $SU(2)$ case, the vector $f$ has only two components and so there
is only one factor in the numerator of the baryon density which thus has
$2(B-1)$ zeros.

\subsection{SPECIFIC FIELDS}

In this section we present the detailed form of harmonic maps which are
used in the construction of the $SU(3)$ skyrmion field ansatze. 

First of all, the $B=1$ case, as discussed in section 5, is
the $SU(2)$ embedded  skyrmion ({\it ie} the hedghog ansatz).
Next we discuss field configurations for $B=2-6$.  
In each case, having found the  map which minimises $\cal{I}$, we solve
numerically (\ref{eq}) and determine the corresponding profile function
$g$.
In Figure 4 we present the energy profiles (as a function of  $r$)
of the resultant skyrmion field configurations. The profiles are given by the 
integrand of (\ref{ene}) where the angular part of the energy, 
contained in $\cal{N}$ and $\cal{I}$, has been integrated. 
In Figure 5 we present the $\theta$ angular dependence of the baryon
densities for $B=2 - 4$ (no $\phi$ dependence).

\begin{figure}[ht]
\epsfxsize=8cm
\hfil\epsffile{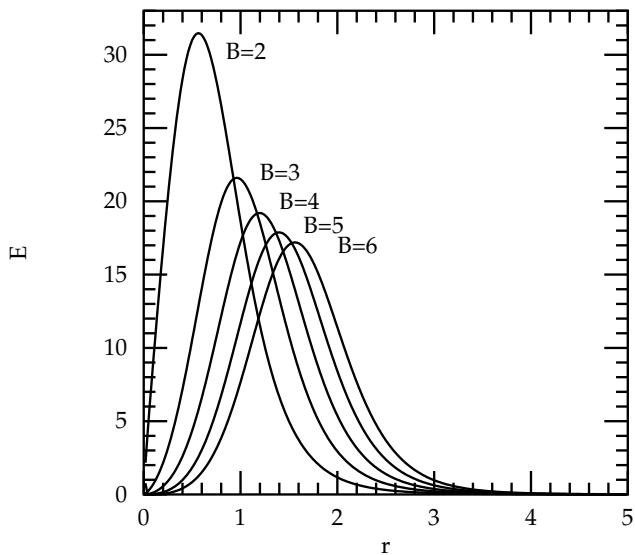}
\caption{Radial energy profiles for $B=2$ to $B=6$.}
\end{figure}

\begin{figure}[ht]
\unitlength1cm
\hfil\begin{picture}(8,8)
\put(3.5,0.35){$\theta$}
\put(3.75,1.9){$B=2$}
\put(3.25,2.7){$B=3$}
\put(5,4){$B=4$}
\put(.5,3){${\cal B}$}
\epsfxsize=7cm
\epsffile{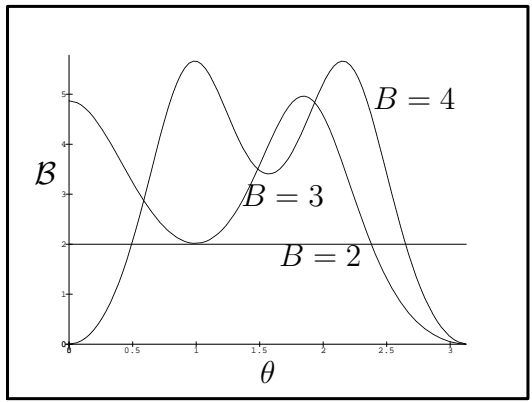}
\end{picture}
\caption{The $\theta$ angular dependence of the baryon density (\ref{bzwg}).}
\end{figure}

$B=2$

Using the ansatz (\ref{genf}), we have minimised $\cal{I}$ numerically 
and have found $f$ to agree with the ansatz presented in section 5, 
{\it ie} to be given by
\be
f=\left( \xi^2, \sqrt{2}\,\xi, \,1
\label{fb2} \right)^t.
\ee
For this field configuration $\cmod{P_+f}/\cmod{f} = 2/(1+\cmod{\xi})^2$
and hence, as shown in Figure 5, the baryon and energy density are
independent of the polar angles on the sphere.
Thus the energy density of the $B=2$ field represents a hollow sphere.

$B=3$

The numerical minimisation of $\cal{I}$ leads to the following expression for 
$f$:
\be
f=\left( \xi^3, \,1.576 \,\xi, \,\sqrt{2}^{\,-1}
\right)^t.
\label{fB3}
\ee  
The baryon density of this configuration is axially symmetric and has the
shape of a torus with a sphere on  top of it. 
In Figures 6a and 6b, we present  plots of surfaces of
constant baryon density for two different values. The values we have chosen are
respectively $0.3$ and $0.7$ times the maximum value of the topological 
density. (In all the graphs that follow, we always express the constant value 
for the curve as a fraction of the maximum density value). Notice that for 
low density value, the three skyrmion configuration has the shape of a
pear, while for higher density values it looks like a ring under a small
ball. 

The energy density has the same symmetry and has a virtually
indistinguishable shape. 
This is also true for all the fields that we will present below. 

Note that as all components of $f$ are monomials, a transformation
$\xi\rightarrow \xi'=\xi e\sp{i\alpha}$, for any $\alpha$ 
({\it ie} a rotation around the $z$-axis), can  be interpreted as an
$SU(3)$ transformation. 
Hence the baryon  density is invariant with respect to such
transformations, {\it ie} it is axially symmetric.

Let us mention that the energy density of our configuration is
remarkably similar to the density of a  $SU(2)$ configuration
corresponding to three skyrmions in a mutually attractive channel \cite{BS2a}
and to the corresponding three  monopole configuration \cite {PSA}.
Given the similarity of  our $SU(3)$ three skyrmion
configuration to the equivalent $SU(2)$
scattering ones as well as to three monopoles one may expect that other
monopole configurations which arise during the scattering process 
might also have their $SU(3)$ analogues. Indeed, as we will see, this 
is the case for our $SU(3)$ four skyrmion configuration.

The baryon density for (\ref{fB3}) does not vanish except when 
$\cmod{\xi}$ is infinite. 
This is the case as the three terms in the numerator
of (\ref{eeee}) do not have common factors; however, as the second term of
(\ref{fB3}) is a polynomial of degree one, the baryon density vanishes
for $\xi=\infty$. 
Indeed, we see in Figures 6a and 6b that the density vanishes on the
negative part of the $z$-axis ($\theta=\pi$).

\begin{figure}
\epsfxsize=14cm
\epsffile{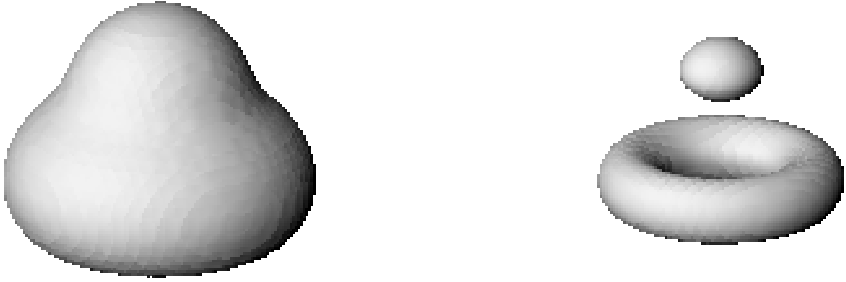}
\caption{Baryon density for B=3:  a: level=0.3  b: level=0.7 }
\end{figure}

\begin{figure}
\epsfxsize=14cm
\epsffile{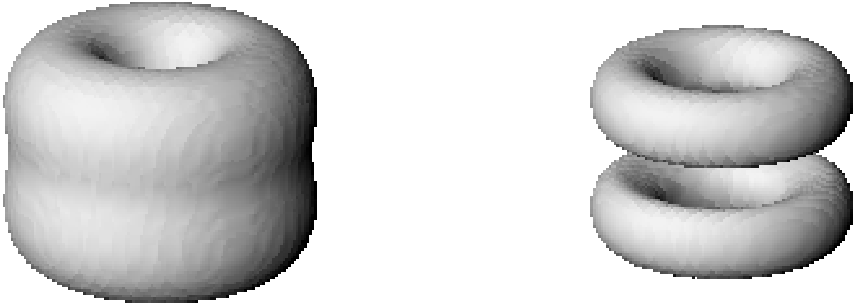}
\caption{Baryon density for B=4:  a: level=0.4  b: level=0.7 }
\end{figure}

$B=4$

For $B=4$ we find 
\be
f=\left(\xi^4, \,2.7191\,\xi^2,\, 1 \right)^t.
\label{fB4}
\ee
This configuration also leads to the energy and baryon
densities that are axially symmetric and they have the shape of two tori 
on top of each other. 
In Figures 7a and 7b, we present  plots of  the surfaces of the
baryon densities at two values.

Once again, the densities corresponding to (\ref{fB4}) are invariant
with respect to a rotation  $\xi \ra  \xi'=\xi e\sp{i\alpha}$.
Note that the baryon density for (\ref{fB3}) does vanish when $\xi$ is
zero or when its modulus $|\xi^2|$ is infinite. 
This happens since the three terms in the numerator of (\ref{eeee})  have
a single common factor at  $\xi = 0$, and the second term of $f$ is a
polynomial of degree two -- implying once again that the baryon density
vanishes when $\xi=\infty$. 
Indeed, this can be seen in Figures 7a and 7b; clearly the density
vanishes
along the $z$-axis ($\theta=0$ and $\theta=\pi$).
Once again, a similar configuration has been observed in the scattering of
four $SU(2)$ monopoles \cite{PSA} and skyrmions in an attractive
channel \cite{BS2a}.

$B=5$

The holomorphic vector for $B=5$ is  given by
\be
f=\left(\xi^5- 2.7\, \xi,\,2\,\xi^4 + 1,\, 9/2\, \xi^3
\right)^t.
\label{fB5}
\ee
Note now that a transformation  $\xi \ra \xi'=i \xi$
({\it ie} a $90^0$ degree rotation around the $z$-axis) corresponds
to a global $SU(3)$ transformation. Hence the densities are invariant
under such transformations. Let us add that the $SU(2)$ embeddings
have very different shapes and symmetries (in fact they are
symmetric under $120^0$ rotations).

It is easy to check that the baryon density corresponding to the
field in (\ref{fB5}) does 
not have any holes. 
Despite  this, one can see holes in Figure 8a and 8b; they correspond to
regions of low, but non-zero, baryon density values. 

Note that by taking $f$ in the form close to (\ref{fB5}), {\it ie}
\be
f=\left(\xi^5+{3C\over D}\, \xi, \,\,
D \,\xi^4 + C, \,\, E\, \xi^3
\right)^t,
\label{fB5a}
\ee
all the three terms in the numerator of (\ref{eeee}) have zeros when 
\be
\xi\sp4={3C\over D},
\label{fB5b}
\ee
which gives four holes in the baryon density. 
So, since our field (\ref{fB5}) is not
very different from (\ref{fB5a}) our densities have minima; corresponding
to  the holes (\ref{fB5b}) partially filled in,
by going from (\ref{fB5a}) to (\ref{fB5}).

\begin{figure}
\epsfxsize=14cm
\epsffile{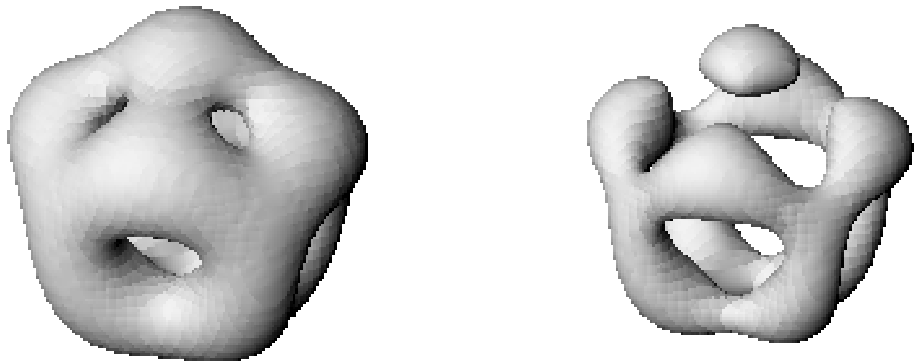}
\caption{Baryon density for B=5:  a: level=0.4  b: level=0.6 }
\end{figure}

\begin{figure}
\epsfxsize=14cm
\epsffile{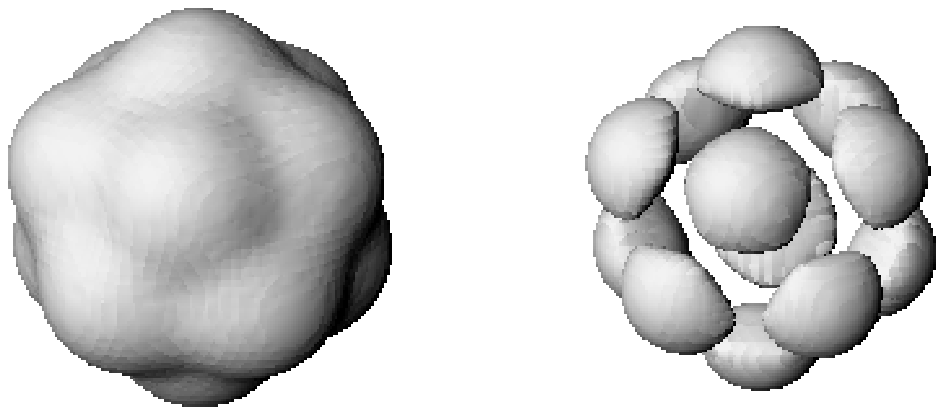}
\caption{Baryon density for B=6:  a: level=0.4  b: level=0.6 }
\end{figure}

$B=6$

The holomorphic vector for $B=6$ is  given by
\be
f=\left(\xi^6+ 3\, \xi, \,\,
1 - 3\, \xi^5, \,\,k\, \xi^3
\right)^t,
\label{fB6}
\ee
where $k$ was found,  numerically, to be $7.06$. 
Once again the baryon density of the field
 (\ref{fB6}) does not have any holes but 
has regions where it is small but non-zero (see Figures 9a and 9b).
These figures show that this configuration has an icosahedral
symmetry and this leads us to the conclusion that, modulo an $SU(3)$ global
transformation, (\ref{fB6}) must be invariant under the following 
transformation \cite{KLEIN}: $\xi\ra \xi'=\xi e\sp{{i2\pi/ 5}}$
({\it ie} a rotation by $72^0$ around the $z$-axis);
$\xi\ra \xi'=- \xi^{-1}$ (which corresponds to
$\theta\ra \pi-\theta$ and $\phi\ra \pi-\phi$) and
$ \xi\ra \xi'=(\xi+b)(b\,\xi+1)^{-1}$ where $b=2 \cos(2 \pi/5) =
(\sqrt5-1)/2$.
This last transformation imposes a condition on $k$ in (\ref{fB6}): 
it is easy to see that the $SU(3)$ transformation on $f$ must be 
of the form $U = R/ \det(R)^{1/3}$ with
\be
R = \left( \begin{array}{lll}
\hs 25 + 15 a &\,\, 10 + 5a &  \hs(150+100) k^{-1} \\[2mm]
\hs 10 + 5 a & 25 + 15 a & -(150 + 100\, a)k^{-1} \\[2mm]
 -k (3 + 2 a) & k (3 + 2 a) &\hs\,\,\,\,\, 15 + 10 a \\[2mm]
\end{array} \right),
\ee 
where $a = -(1 + \sqrt5)/2$.
Imposing the condition that the rows and columns of $R$ are 
orthogonal to each other implies that $k = \sqrt50 \approx 7.071$ which is 
within the precision of our numerical minimisation programme.

Having presented our field configurations we can now discuss
some of their properties and compare them with the $SU(2)$ 
embeddings.

First of all, for $B=1$ we have only the $SU(2)$ embedding. Its energy
and baryon density is in the shape of a ball.
For $B>1$ our field configurations are different from the 
$SU(2)$ embeddings. 
However,  the densities of the baryon densities for $B\le4$ are all
axially symmetric (see Figure 5).
The $B=2$ configuration is radially symmetric and the
baryon density corresponds to a shell (in contrast to the toroidal $SU(2)$
one); the $B=3$ configuration corresponds
to a single skyrmion located around the north pole of the $S^2$ sphere
and the other two are below the equator (spread out to form a charge two 
torus-like structure), while the $B=4$ configuration consists of four
baryons which are in the shape of two partially
overlapping tori close to the equator of the sphere.
The fields for $B>4$ are more complicated, their baryon densities
have fewer symmetries as seen from our figures. The baryon and
energy densities for the case of $B=5$ resemble a structure
consisting of two deformed tori, close to the equator, with an additional ball
at the north pole of the angular sphere while for $B=6$ they 
form a structure which is icosahedrally symmetric. 

These shapes are very different from what was seen for $SU(2)$ fields and,
as we have discussed  above, they have also different symmetries.

In Figure 4, where  we have plotted the energy profile functions for baryon 
numbers from two to six we note that the effective size of the baryons 
increases with the 
increasing baryon number -- this is reflected in the  shift to the right
of the profile functions. 

In the following table we present the energy values of the resulting
Skyrme fields. 
All the numerical values of the energies are given in units of $B$  
and hence are close to unity. 
These values are then compared with the  $SU(2)$ skyrmion embeddings
obtained using rational maps in \cite{HMS}.
We see that both field configurations have similar values of energy, although
the energies of the embeddings are marginally lower.

\begin{table}[htbp]
\begin{center}
\begin{tabular}{c|c|c|c|c}
$B$ & ${\cal I} \,(SU(3))$ & $SU(3)$ En/Sk (Ansatz) & $SU(2)$ 
 En/Sk (Ansatz)& $SU(2)$ En/Sk (Numerical)\\
\hline
1 &1& 1.232 &1.232 & 1.232\\
2 &4& 1.222 &1.208 & 1.171\\
3 &10.65356 & 1.215&1.184 & 1.143\\
4 &18.04501 &1.184 &1.137 & 1.116\\
5 & 27.26 & 1.164 & 1.147 & 1.116\\
6 & 37.33 & 1.1458 & 1.137 & 1.109\\
\end{tabular}
\end{center}
\caption{Energy of $SU(3)$  harmonic ansatz compared to the energy of the 
$SU(2)$ harmonic ansatz and the energy of the $SU(2)$ solutions obtained 
numerically \cite{HMS}.}
\end{table}

\section{CONCLUSIONS}

In this paper, we have discussed various static field
configurations of the $SU(N)$ Skyrme model.
We have shown that, in addition to the obvious embeddings,
any solution of the $SU(2)$ model generates a solution of the 
$SU(4)$ model. 
Unfortunately, this solution is topologically trivial ({\it ie} its baryon
density vanishes identically) and its energy is four times the original
$SU(2)$ solution.

Next we have generalised the harmonic map ansatz of Houghton at al \cite{HMS}
and showed that this ansatz has allowed us to find another 
exact solution, this time of,  the $SU(3)$
model.
The baryon number of this solution is also zero and its energy density is
radially symmetric. 
However, its total energy  is less than four in topological
units and we have argued that it represents a bound state of
two skyrmions and two anti-skyrmions.

Using our generalisation of the harmonic map ansatz we
have then presented topologically nontrivial field configurations
of the $SU(N)$ Skyrme model with radially symmetric energy densities. 
They correspond to $B=N-1$ skyrmions in $SU(N)$ models. 
In the massless case their energies have turned out also to be above
those of the $SU(2)$ embeddings.
However, when mass is added to the model, for sufficiently large
masses, their energies can be be lower than the energies of the
embeddings. 

We have also looked at various field configurations of the $SU(3)$ model.
The energy and baryon densities of these $SU(3)$ fields exhibit shell-like
structures; in all cases, except for $B=1$, they are different from the
corresponding structures seen in the $SU(2)$ model and more
symmetrical. 
Their energies are slightly higher but comparable to those of the
embeddings. 
Their different symmetry properties suggest to us that although these
embeddings have higher energies they may be  reflections of real states of
the model showing that the $SU(3)$ model can have many interesting
solutions. 
To see whether this expectation is correct one has to perform numerical
simulations - this so far has not been done.

Finally, our projector ansatz suggests that one might try to construct 
further ansatze involving two or more projectors. Such  ansatze will then 
depend on more that one profile function. 
This topic is currently under investigation. 

\section{ACKNOWLEDGMENTS}

The authors would like to thank V. Kopeliovich and  J. Garraham
for their help in finding a numerical mistake in the earlier report of
this work (in the form of two papers).\\
We also thank C. J. Houghton for his interest ans correspondence about the 
symmetries of our field configurations.


\begin{thebibliography}{99}
\bibliographystyle{plain}
\bibitem{HMS}  
C. J. Houghton, N. S. Manton and P. M. Sutcliffe, Nucl. Phys. B {\bf 510},
507 (1998).    
\bibitem{Skyr1}
T. H. R. Skyrme, Nucl. Phys. {\bf 31}, 556 (1962).
\bibitem{Wit}
E. Witten, Nucl. Phys. B {\bf 223}, 422 (1983).
\bibitem{M}
N. S. Manton, Phys. Lett. B {\bf 192}, 177 (1987).
\bibitem{BS}
R. A. Battye and P. M Sutcliffe, Phys. Rev. Lett. {\bf 79}, 363 (1997).
\bibitem{Skyr}
T. H. R. Skyrme, Proc. R. Soc. A {\bf 260}, 127 (1961).
\bibitem{BBLRS}
A. P. Balachandran, A. Barducci, F. Lizzi, V. G. J. Rodgers and A. Stern,
Phys. Rev. Lett. {\bf 52}, 887 (1984). 
\bibitem{Kop} 
V.B. Kopeliovich, B.E. Schwesinger and B.E. Stern, JETP Lett. {\bf 62}, 
185-90 (1995).
\bibitem{Zak} 
W. J. Zakrzewski,  Low dimensional sigma models (IOP, 1989).
\bibitem{DinZak}
A. Din and W. J. Zakrzewski, Nucl. Phys. {\bf B 174}, 397 (1980).
\bibitem{Dada}
A. D'Adda, P. Di Vecchia and M. Luscher, Nucl. Phys. B {\bf  146}, 63
(1980).
\bibitem{BS2a}
R. A. Battye and P. M Sutcliffe, in preparation.
\bibitem{PSA}
P. M. Sutcliffe, Nucl. Phys. {\bf 505}, 517 (1997).
\bibitem{KLEIN}
F. Klein, The Icosahedron, Dover Publications, (1956).
\end{thebibliography}
\end{document}